\documentclass[aps,prd,preprint,superscriptaddress]{revtex4}

\usepackage{graphicx}
\usepackage{amsmath}
\usepackage{amsfonts}

\usepackage{graphicx}
\usepackage{dcolumn}
\usepackage{bm}
\usepackage{amssymb}

\begin{document}

\title{Formation of metallic magnetic clusters in a Kondo-lattice metal:  Evidence from an optical study}
\author{N.~N. Kovaleva}
\email{N.Kovaleva@lboro.ac.uk,NatalyN.Kovaleva@googlemail.com}
\affiliation{Department of Physics, Loughborough University, LE11 3TU Loughborough, United Kingdom}
\affiliation{Institute of Physics, ASCR, 18221 Prague, Czech Republic}
\author{K. I. Kugel}
\affiliation{Institute for Theoretical and Applied Electrodynamics, Russian
Academy of Sciences, 125412 Moscow, Russia}
\author{A. V. Bazhenov}
\author{\mbox{T. N. Fursova}}
\affiliation{Institute for Solid State Physics, Russian Academy of Sciences,
142432 Chernogolovka, Russia}
\author{W.~L\"oser}
\affiliation{Leibniz Institut f\"ur Festk\"orper- und Werkstoffforschung Dresden, D-01171 Dresden, Germany}
\author{Y.~Xu}
\affiliation{Leibniz Institut f\"ur Festk\"orper- und Werkstoffforschung Dresden, D-01171 Dresden, Germany}
\affiliation{State Key Laboratory of Solidification Processing, Northwestern Polytechnical University, Xi'an 710072, P.~R. China}
\author{G.~Behr}
\altaffiliation{Deceased.}
\affiliation{Leibniz Institut f\"ur Festk\"orper- und Werkstoffforschung Dresden, D-01171 Dresden, Germany}
\author{F. V. Kusmartsev}
\affiliation{Department of Physics, Loughborough University,
LE11 3TU Loughborough, United Kingdom}

\date{\today}

\pacs{71.20.Lp 71.27.+a 78.20.-e}

\maketitle

{\bf Magnetic materials are usually divided into two classes: those with localised magnetic moments, and those with itinerant charge carriers. We present a comprehensive experimental (spectroscopic ellipsomerty) and theoretical study to demonstrate that these two types of magnetism do not only coexist but complement each other in the Kondo-lattice metal, Tb$_2$PdSi$_3$. In this material the itinerant charge carriers interact with large localised magnetic moments of Tb(4f) states, forming complex magnetic lattices at low temperatures, which we associate with self-organisation of magnetic clusters. The formation  of magnetic clusters results in low-energy optical spectral weight shifts, which correspond to opening of the pseudogap in the conduction band of the itinerant charge carriers and development of the low- and high-spin intersite electronic transitions. This phenomenon, driven by self-trapping of electrons by magnetic fluctuations, could be common in correlated metals, including besides Kondo-lattice metals, Fe-based and cuprate superconductors.}

\vspace{1cm}
The quantum mechanics' Pauli principle, also known as the exclusion principle, postulates that a single orbital may accommodate no more than two electrons and, if it is doubly occupied, the electron spins must be paired. This principal
is implicated in the electronic and magnetic properties of solids as the
so-called {\it exchange effect}. In the Stoner model of ferromagnetism of weakly-correlated itinerant electrons (the form observed in the elemental metals iron, cobalt and nickel) electron bands can spontaneously split into up and down spins. This happens if the relative gain in an exchange interaction is larger than the loss in kinetic energy \cite{Stoner}. On the other hand,
in the Heisenberg model of localised magnetism in magnetic insulators (the form observed in compounds LaMnO$_3$, KCuF$_3$, V$_2$O$_3$, and many others) the electrons, involved in the exchange coupling, are localised by strong electronic correlations when the large on-site Coulomb repulsion $U$ \cite{zaanen,arima,kovaleva_jetp}
opens a well-defined gap for charge carriers; and an adequate description of the ground state properties of these systems in terms of the superexchange spin-orbital models has been suggested
\cite{goodenough,kanamori,khomskii,feiner,oles,kovaleva_lmo_prl,kovaleva_lmo_prb,qazilbash,basov}.

\begin{figure}[b]
\includegraphics[width=12.5cm]{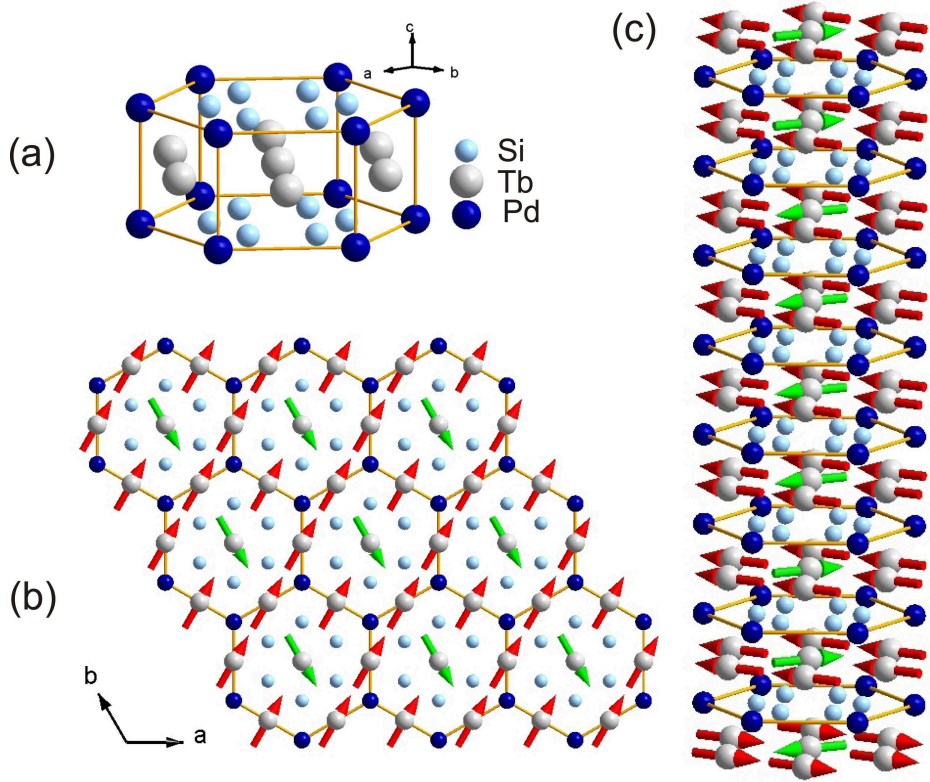}
\caption{{\bf Structure and long-range magnetic order in Tb$_2$PdSi$_3$.} (a) A simplified unit cell of Tb$_2$PdSi$_3$, neglecting the {\bf c}-axis superstructure. (b,c) {\bf ab}-plane and {\bf c}-axis magnetic ordering patterns
according to the results of neutron scattering study \cite{FrontzekThesis}. \vspace{-1.5em}}
\label{Fig1}
\end{figure}

However, recently synthesised Pd-based ternary compounds of the form $R_2$PdSi$_3$, where $R$ is a rare earth atom \cite{CrystalGrowth}, have been found to have
pronounced anomalies in resistivity, magnetization and susceptibility, associated in literature  with their unusual or ``novel''
magnetic properties~\cite{FrontzekKreyssig06,FrontzekKreyssig07,Paulose,FrontzekThesis,Chevalier,Szytula99}.
These compounds  descend from $R$Si$_2$ compounds with the layered AlB$_2$-type
crystal structure, and have quasi-hexagonal symmetry [see Fig.\,\ref{Fig1}\,(a)].
The Tb rare-earth compound  reveals the most intriguing regime of quasi-one-dimensional magnetism and anisotropic ``spin-glass''-like behaviour
(see Refs. \cite{FrontzekKreyssig06,FrontzekKreyssig07,Paulose,FrontzekThesis}
and references therein). In this compound the states at the Fermi level  ($E_{\rm F}$) were found to be dominated by the highly conducting itinerant electrons of the Tb 5d$^1$ orbital, whereas the Tb 4f states responsible for the large magnetic moment (of 9.68 $\mu_B$, comparable to that of the free-ion Tb$^{3+}$ value of 9.72 $\mu_B$)  were found to be localised deep below $E_{\rm F}$ near 8\,eV \cite{ChaikaIonov01}. In this paper we demonstrate
that the unusual magnetic properties of Tb$_2$PdSi$_3$ are the result of
complementary effects of itinerant and localised magnetism  in the Kondo-lattice metallic state. The itinerant charge carriers interact with large localised magnetic moments of the Tb 4f states, forming regular lattices of self-organised ferromagnetic clusters below the N\'eel temperature $T_{\rm N}$ = 23.6 K.

To elucidate the nature of electronic instabilities in the Kondo-lattice metal Tb$_2$PdSi$_3$, and the resulting  unusual magnetic ordering phenomena, we used a comprehensive spectroscopic ellipsometry approach, which was successfully applied in our earlier studies  of electronic correlations in the magnetic insulators \cite{kovaleva_lmo_prl,kovaleva_lmo_prb,kovaleva_yto}.

So far the properties of the metallic Kondo-lattice compound  Tb$_2$PdSi$_3$
have been discussed in terms of quasi-one-dimensional magnetism at high temperatures
(near 55 K) and anisotropic ``spin-glass''-like behaviour at low temperatures
($<$ 10 K) \cite{FrontzekKreyssig06,FrontzekKreyssig07,Paulose,FrontzekThesis}.  In particular, the broad peak in magnetic susceptibility near 55 K has been attributed to the magnetic correlations arising in one-dimensional spin chains, which have been intensively studied since the seminal paper by Bonner and Fischer \cite{Bonner1964}. The peak exhibits strong anisotropy in the paramagnetic state, also suggesting some similarities with quasi-one-dimensional systems. All the results convincingly demonstrate that some kind of long-range magnetic ordering sets in below $T_{\rm N}=23.6$ K \cite{FrontzekKreyssig06,FrontzekKreyssig07,Paulose,FrontzekThesis}. However, the nature of the magnetism appears to be very complex, with strong
evidence of ferromagnetic correlations and a complex
ferrimagnetic type order in the  basal plane, and a kind
of anti- and ferromagnetic correlations perpendicular to the basal
plane, along the {\bf c} axis \cite{FrontzekThesis}.

Neutron diffraction studies of this compound in zero magnetic field revealed both the long- and short-range order \cite{FrontzekKreyssig06,FrontzekKreyssig07,FrontzekThesis}.
The long-range order (LRO) below $T_{\rm N}=23.6$ K may be described as follows.
The magnetic unit cell of  Tb$_2$PdSi$_3$  is four 
times as big in the basal plane, contains four magnetic Tb$^{3+}$ ions
(at the $3f$ and $1a$ site positions of $P6/mmm$ hexagonal symmetry), and
is sixteen times as big along the {\bf c} direction. While the Tb moments are lying within the basal plane, as illustrated by Fig.\,\ref{Fig1}\,(b), the most intriguing is the spin sequence of the  magnetic moments associated with the individual Tb$^{3+}$ ions along the {\bf c} direction.
The neutron diffraction study \cite{FrontzekThesis} suggests that for all $3f$ positions there is a spin-sequence of eight positive orientation of the Tb moments ``$+$'', followed by their eight negative  orientation ``$-$''.
On the other hand, the magnetic moments at $1a$ positions generate a different spin-sequence, consisting of the alternating orientations ``$++ - - - - ++ - - ++++ - -$'', as schematically presented in Fig.\,\ref{Fig1}\,(c).
The low-temperature short-range correlations (SRC) are suggested to arise from frustrated magnetic moments within the LRO, and are associated with the ``spin-glass''-like behaviour detected in the magnetic susceptibility measurements~\cite{FrontzekKreyssig06,FrontzekKreyssig07,Paulose,FrontzekThesis}.
We argue that the complex ``exotic'' phases observed in this compound in an external magnetic field and the low-temperature ``glassy'' behavior are
associated with self-organisation of magnetic clusters, which can merge into string-like droplets or stripy microdomains, may be ordered or disordered.

Our main results are as follows. After a thorough study of the optical response of the Kondo-lattice metal Tb$_2$PdSi$_3$ at low energies, we discovered anisotropic anomalous optical spectral weight (SW) changes across the  N\'eel temperature $T_{\rm N}=23.6$ K, involving the free charge carrier (Drude) resonance and two midinfrared (MIR) optical bands at 0.2\,eV and 0.6\,eV.
We found that a vast amount of the optical SW of the free charge carriers
(of the Drude peak) is shifted to the optical band at
0.2\,eV. The observed anomalous behaviour in the low-energy optical conductivity spectra below 0.2 eV can be associated with the opening of a pseudogap in the conduction band and reconstruction of the Fermi surface. These anomalies
are driven by electronic correlations related to the free electron instability against self-trapping in the metallic ferromagnetic cluster, rather than band structure effect. Localisation of electrons in these magnetic clusters
leads to development of the high-spin (HS) and low-spin (LS) inter-site electronic
transitions, which were determined by their anomalous behavior in our comprehensive temperature study around $T_{\rm N}$ at 0.2 and 0.6\,eV, respectively.

Finally, we note that in the optical conductivity spectra of the
Fe-based superconductors, namely BaFe$_2$As$_2$ and
Fe-chalcogenides, optical bands at around 0.2\,eV and 0.6\,eV,
which evolve with electron- or hole- doping have been identified~\cite{Hu,Wu,Wang}. In addition, the midinfrared features at 0.2\,eV and 0.6\,eV are well-known
and, according to numerous studies,  are strongly pronounced in the
optical conductivity spectra of all high-$T_c$ cuprates, whereas
they are absent in their dielectric phases~\cite{Bazhenov1,Bazhenov2}.
We can argue that electronic correlations lead here to the energy lowering due to self-trapping of charge carriers by magnetic fluctuations, thus
giving rise to nanoscale inhomogeneities with a local magnetic
ordering. The existence of these nanoscale inhomogeneities must
manifest itself in the midinfrared optical conductivity as the
high-spin (at around 0.2\,eV) and low-spin (at around 0.6\,eV) optical excitations.

\begin{figure}[b]
\includegraphics[width=12.5cm]{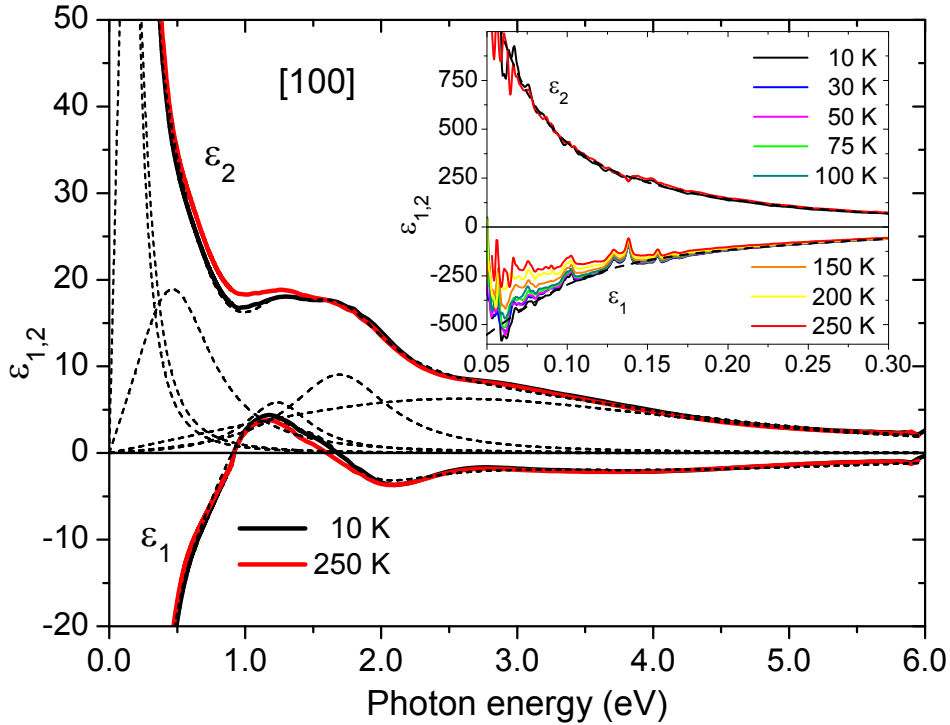}
\caption{{\bf Wide-range optical response of Tb$_2$PdSi$_3$.} Low-temperature
(10\,K) and high-temperature (250\,K) dielectric functions of Tb$_2$PdSi$_3$ in [100] polarisation. Constituents of the classical dispersion ana\-ly\-sis of the complex dielectric response at 10\,K in terms of the Drude-Lorentz model \cite{Allen} are shown by dashed curves. Inset: Zoom of the low-energy Drude response.\vspace{-1.5em}}
\label{Fig2}
\end{figure}

\begin{figure*}[t]\hspace{-1ex}
\includegraphics[width=17.5cm]{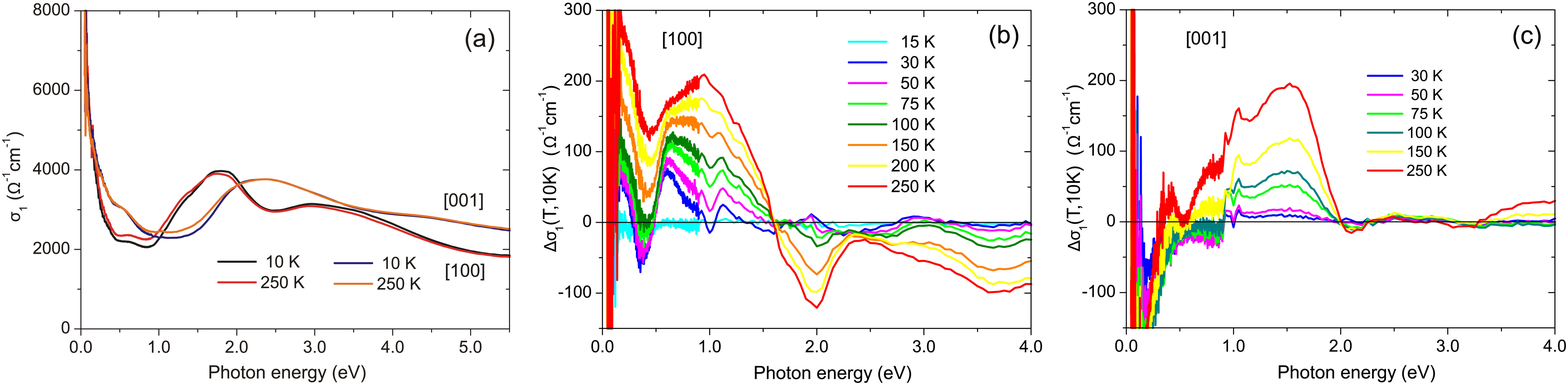}
\caption{{\bf Temperature-dependent anisotropic optical conductivity.}
(a) Low-temperature (10\,K) and high-temperature (250\,K) optical-conductivity spectra, $\sigma_{\kern-0.5pt1}(\omega)$, and their anisotropy in [100] and [001] polarisations. (b,\,c) Details of the temperature dependence of $\sigma\!_1(\omega,T)$ in the difference spectra, $\sigma\!_1(\omega,T)-\sigma\!_1(\omega,10\,{\rm K})$, in [100] and [001] polarisations, respectively.\vspace{-1.2em}}
\label{Fig3}
\end{figure*}

\vspace{0.3cm}
\hspace{-1.2em}{\large \bf Results}\\
\hspace{-1em}{\bf Wide-range optical response of Tb$_2$PdSi$_3$}\\
Here we report the results of a spectroscopic ellipsometry study of a
wide-range anisotropic dielectric response of high-quality single crystals
of Tb$_2$PdSi$_3$. The available data on the optical properties of $R_2$PdSi$_3$ ternary silicides and their parent compounds $R$Si$_2$ is limited, to the best of our knowledge, to a single study of the room-temperature optical reflectivity  in GdSi$_2$ and ErSi$_2$ polycrystalline films \cite{Guizzetti}.
The low- and high-temperature spectra of the real and imaginary parts of the dielectric function, $\varepsilon(\omega)=\varepsilon_1(\omega)+{\rm i}\kern.5pt\varepsilon_2(\omega)$, in [100] polarisation are shown in Fig.\,\ref{Fig2}. They demonstrate the main features of the $\bf a$-axis dielectric response of Tb$_2$PdSi$_3$. At low energies, the dielectric function is dominated by a Drude-type response due to transitions within the conduction
band, characterized by the free charge carrier scattering rate, $\gamma_{\rm D}$, and the plasma frequency, $\omega_{\rm p}$. The zoomed-in low-energy part of the dielectric function is shown in the inset, where we also show $\varepsilon_1(\omega)$ at intermediate temperatures between 10 and 250 K. At higher energies, broad optical bands associated with interband transitions can be identified by the resonance and antiresonance features that appear at identical energies in $\varepsilon_2(\omega)$ and $\varepsilon_1(\omega)$, respectively, obeying the Kramers-Kronig relations. The spectra can be well described in terms of the Drude-Lorentz model \cite{Allen}, with the results of the dispersion analysis summarised in Fig.\,\ref{Fig2} and in Table~1 (see Table~1 in the Supplementary Information to this paper). The optical response in the low-energy region is composed of a Drude peak and two bands centered in the MIR region at $\omega_1\approx0.2$\,eV and at $\omega_2\approx0.6$\,eV. At higher energies, we identify at least three interband transitions centered at 1.3, 1.7 and 3.4\,eV. The $\bf c$-axis dielectric function exhibits similar features in the optical response (not shown), however, compared to the $\bf a$-axis spectra, we notice pronounced anisotropy in the range of the interband optical transitions, which appear at 2.2, 2.7 and 5.2\,eV. The  LDA calculations reproduce the main features of the optical response of Tb$_2$PdSi$_3$ and its anisotropy reasonably well.

Fig.\,\ref{Fig3}\,(a) shows the related $\bf a$- and $\bf c$-polarised optical-conductivity spectra, $\sigma\!_1(\omega)=\omega\varepsilon_2(\omega)/(4\pi)$, in which the MIR optical band at $\omega_2\approx0.6$\,eV and the higher-energy bands are more pronounced than in $\varepsilon_2(\omega)$. Fig.\,\ref{Fig3}\,(a) also reflects changes in the optical-conductivity spectra at elevated temperatures. In the $\bf a$-polarized spectra, the high-energy optical bands at 1.3, 1.7, and 3.4\,eV broaden, while their intensities decrease. The peak position of the strongly pronounced optical band at 1.7\,eV exhibits a noticeable
shift, and as a result, an isosbestic point appears in the temperature-dependent optical-conductivity spectra at 1.5\,eV, associated with a displacement of the optical SW to lower photon energies. In the $\bf c$-axis spectra, the broadening of the high-energy optical bands at 2.2 and 2.7\,eV explains notable changes at photon energies between 1 and 2\,eV.

To get a better insight into the temperature-dependence of the
$\bf a$- and $\bf c$-polarised optical conductivity spectra, in
Fig.\,\ref{Fig3}\,(b,\,c) we plot the difference between
$\sigma\!_1(\omega,\,T)$ and the respective low-temperature
spectrum, $\sigma\!_1(\omega,\,10\,{\rm K})$. It can be seen here
that the major temperature effects within the high-energy optical
bands, discussed above, occur between 100 and 250\,K and,
therefore, can be associated with lattice anharmonicity effects.
However, qualitatively different changes occur in the $\bf a$ and
$\bf c$-polarised optical-conductivity spectra at low energies.
Indeed, here the Drude response expands to higher photon energies
with increasing temperature due to an increase in $\gamma_{\rm
D}$, together with simultaneous changes in the intensity and width
of the MIR optical bands at 0.2 and 0.6\,eV. Qualitatively
different changes at low energies in the $\bf a$- and $\bf
c$-polarized spectra indicate that  the temperature changes for
these competing effects are polarisation
dependent.

\vspace{0.3cm}
\hspace{-1em}{\bf Anomalies induced by magnetic ordering}\\
The anomalous behavior of the optical bands at $\omega_1\approx0.2$\,eV
and $\omega_2\approx0.6$\,eV at the low temperature also becomes evident close to the magnetic phase transition, near $T_{\rm N}=23.6$\,K. Here the difference spectra exhibit a dip, followed by a maximum around 0.6\,eV. This feature has an asymmetric shape, with an extended tail at higher photon energies, where the second weaker maximum shows up around 1.2\,eV. The latter exhibits a similarly anomalous behavior near $T_{\rm N}$. One can compare the rearrangement of the SW due to these low-temperature anomalies with the anharmonicity coming into effect at elevated temperatures.

\begin{figure}[t]
\includegraphics*[width=14.0cm]{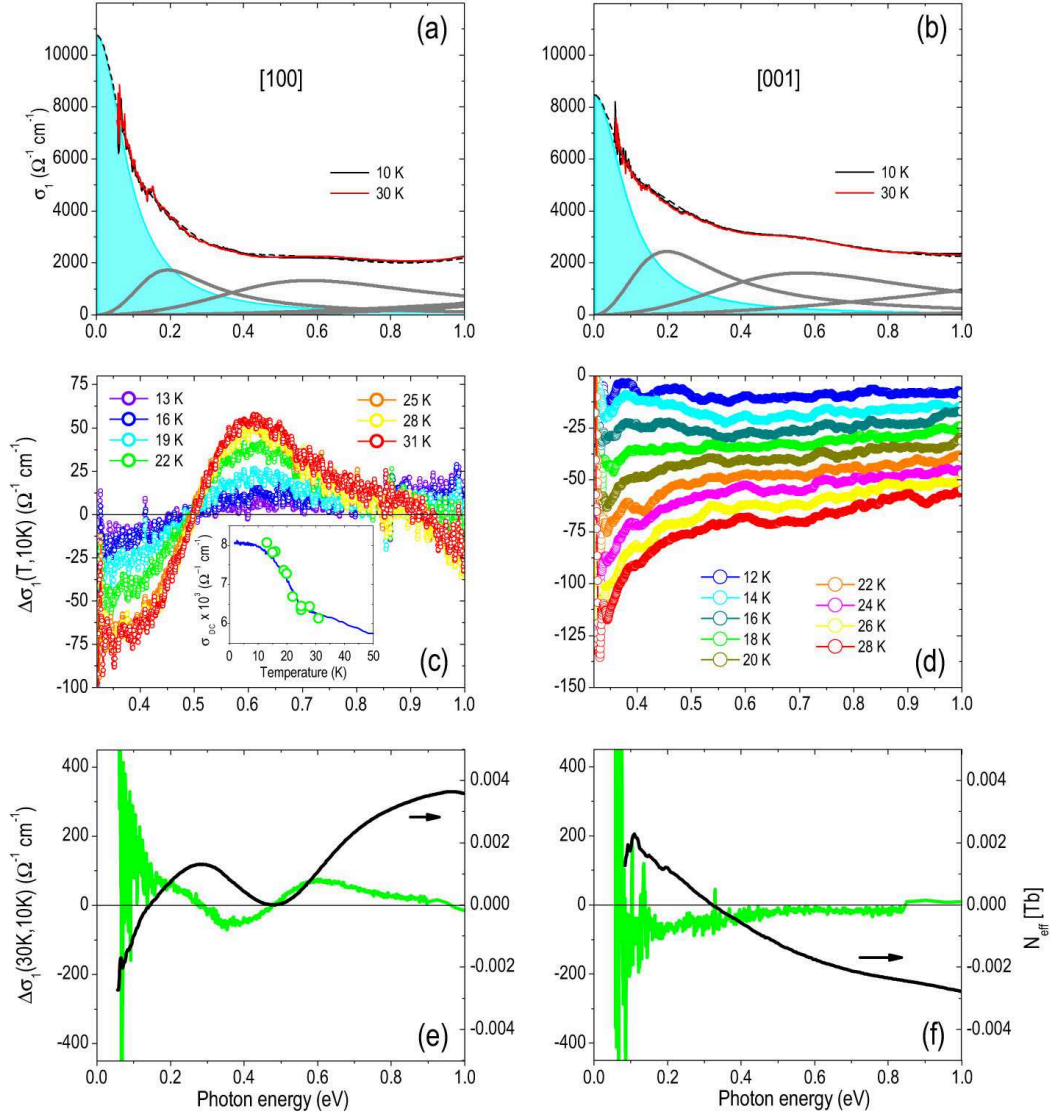}
\caption{{\bf Anomalies in optical conductivity induced by magnetic ordering.}
Left and right panels correspond to [100] and [001] polarisations, respectively. (a,\,b) Low-temperature (10 and 30\,K) optical-conductivity spectra, $\sigma_{\kern-0.5pt1}(\omega)$, the contribution from the Drude term (shaded area), and the Lorentz optical bands below 1\,eV. (c,\,d) Critical behavior of the MIR optical conductivity across $T_{\rm N}$. Inset: $(1-{\rm const}\times I_{\rm 0.6\,eV}^{\rm max})$ superimposed on a linear scale with the $\bf a$-axis dc conductivity (see the Supplementary Information to this paper) \cite{Majumdar}. (e,\,f) ${\scriptstyle\rm\Delta}\sigma\!_1$(30\,K,\,10\,K)
(green curve) and the associated SW changes (black curve).\vspace{-1.2em}}
\label{Fig4}
\end{figure}

Figure\,\ref{Fig4}\,(a,\,b) shows relatively weak changes in the $\bf a$- and $\bf c$-polarised optical-conductivity spectra at low photon energies ($\hbar \omega$\,=\,0.05--1.0\,eV) around $T_{\rm N}$ and presents the results of the Drude-Lorentz analysis at 10\,K. Contributions from the Drude term and from the MIR optical bands at 0.2 and 0.6\,eV can be clearly seen here. Panels (c) and (d) of the same figure emphasise the anomalous temperature
dependencies of the anisotropic MIR optical conductivity near $T_{\rm N}$. The inset of Fig.\,\ref{Fig4}\,(c) shows critical behavior of the peak intensity, $I_{\rm 0.6\,eV}^{\rm max}$, of the 0.6\,eV optical band, such that $(1-{\rm const} \times I_{\rm 0.6\,eV}^{\rm max}$) follows trends in  the dc transport anomalies in the $\bf a$-axis spectra near $T_{\rm N}$ (see the Supplementary Information to this paper) \cite{Majumdar}. And finally, we analyze the difference, $\sigma\!_1(\omega,30\,\text{K})-\sigma\!_1(\omega,10\,\text{K})$, of the spectra measured immediately above and well below $T_{\rm N}$, which exhibits clear anisotropy in the corresponding $\bf a$- and $\bf c$-polarised spectra [Fig.\,\ref{Fig4}\,(e,\,f)].

According to our dispersion analysis of the $\bf a$-axis spectra (see Table~1 in the Supplementary Information to this paper), in addition to the changes associated with the Drude response, the oscillator strength of the optical band at 0.2\,eV decreases, whereas the one at 0.6\,eV increases above $T_{\rm N}$. Low-energy downturn in this difference indicates that the crossing point for the SW transfer between the ``head'' of the Drude response and the higher-energy response is located near 0.06\,eV. Above 0.06\,eV, the changes due to the Drude response compete with the opposite changes associated with the MIR optical band at 0.2\,eV. The cumulative changes of the Drude response and the involved MIR optical bands result in the creation of zero crossing points around 0.27\,eV and 0.47\,eV.

In order to quantify the SW involved in these anomalies, we have integrated the associated optical SW increment as ${\scriptstyle\rm\Delta}\text{SW}(\omega,{\scriptstyle\rm\Delta}T)=\int_{0}^{\omega}{\scriptstyle\rm\Delta}\sigma\!_1(\omega',{\scriptstyle\rm\Delta}T)\,{\rm d}\omega'$. In this investigation the optical conductivity changes of the Drude response across $T_{\rm N}$, $\Delta \sigma^{\rm D}_1(\omega,30\,{\rm K},\,10\,{\rm K})$, resulting from our dispersion analysis (see Table~1 in the Supplementary Information to this paper), are incorporated. For more accurate estimate of the associated energy scale, where the total optical SW obeys the f-sum rule \cite{Dressel}, an additional optical study at lower frequencies, which enables consistency in the dc limit $\sigma_{\kern-0.5pt1}(\omega\rightarrow0,T)$, is required. The resulting SW changes are presented in Fig.\,\ref{Fig4}\,(e) in terms of the effective number of charge carriers per Tb atom, ${\scriptstyle\rm\Delta}N_{\rm eff}=2V\!m\,{\scriptstyle\rm\Delta}{\rm SW}/\pi e^2$, where $m$ and $e$ are the free-electron mass and charge, and $V=\sqrt{3}a^2c/2$ is the unit cell volume corresponding to one Tb atom, expressed through the low-temperature lattice parameters $a=4.0643$\,\AA\ and $c=4.0502$\,\AA\ \cite{Szytula99}. Alternatively, we estimate the optical SW associated with the contribution of charge carriers as $N^{\rm D}_{\rm eff}=\frac{2Vm}{\pi e^2} \frac{(2\pi c)^2\omega_{\rm p}^2}{8}$, which with $\omega_{\rm p}=2.703$\,eV gives $N^{\rm D,\bf ab}_{\rm eff}(10\,{\rm K})=0.308$ and with $\omega_{\rm p}=2.750$\,eV gives $N^{\rm D,\bf ab}_{\rm eff}(30\,{\rm K})=0.319$ (see Table~1 in the Supplementary Information to this paper). Hence, the SW changes of the Drude response across $T_{\rm N}$, ${\scriptstyle\rm\Delta}N^{\rm D,\bf ab}_{\rm eff}(30\,{\rm K},\,10\,{\rm K})=0.011$, amount to 3.6\% of the total low-temperature charge carrier SW, $N^{\rm D,\bf ab}_{\rm eff}(10\,{\rm K})$. As evident from Fig.\,\ref{Fig4}\,(e), these SW changes are nearly compensated around the zero-crossing point at 0.47 eV, by the opposite changes of the SW of the optical band at 0.2 eV (which are in a good quantitative agreement with the SW amplitude around 0.2\,eV  across the spin-density-wave (SDW) transition at $T_{SDW}=200$ K in the iron arsenide SrFe$_2$As$_2$ \cite{charnukha}).
Therefore, the total SW gain across $T_{\rm N}$ accumulated within the energy window $\hbar\omega$\,$\lesssim$\,1\,eV in the $\bf ab$ plane, ${\scriptstyle\rm\Delta}N^{\rm\bf ab}_{\rm eff}(30\,{\rm K},\,10\,{\rm K})$\,$\approx$\,0.0036 [see Fig.\,\ref{Fig4}\,(e)], amounts to 1.2\% of the total charge carrier SW, $N^{\rm D,\bf ab}_{\rm eff}(10\,{\rm K})$, and can be attributed to the SW gain of the optical band at 0.6 eV.

In the $\bf c$-axis optical-conductivity spectra, in addition to the changes associated with the Drude response, the oscillator strength of the optical band at 0.2\,eV decreases above $T_{\rm N}$, while the one at 0.6 eV exhibits weak temperature dependence (see Table~1 in the Supplementary Information to this paper).
Our analysis represented in Fig.\,\ref{Fig4}\,(f) shows that the SW changes of the Drude response across $T_{\rm N}$ are nearly compensated at around 0.23 eV, by the opposite changes of the SW of the optical band at 0.2 eV. The total SW loss across $T_{\rm N}$ within the energy window $\hbar\omega$\,$\lesssim$\,1\,eV in the $\bf c$ axis amounts to ${\scriptstyle\rm\Delta}N^{\rm \bf c}_{\rm eff}(30\,{\rm K},\,10\,{\rm K})$\,$\approx-$\,0.0028 [see Fig.\,\ref{Fig4}\,(f)].

\vspace{0.3cm}
\hspace{-1.2em}{\large \bf Discussion}\\
We found that in the temperature  range 10 K $\leq T \leq
T_{\rm N}$, anomalous changes of the optical SW in Tb$_2$PdSi$_3$
occur below 1\,eV, including the Drude-type resonance and the MIR
optical bands at 0.2 and 0.6\,eV. We attribute these anomalies to electronic
instabilities (many-body effects) caused by the nature of magnetism inherent in this Kondo-lattice metal, which exhibits quasi-one-dimensional characteristics
at higher temperatures and spin-glass-like behaviour at lower temperatures
\cite{FrontzekKreyssig06,FrontzekKreyssig07,Paulose,FrontzekThesis}. The complex long-range magnetic order below $T_{\rm N}$ is due to  ferromagnetic
correlations in the ferrimagnetic-type arrangements of the large Tb moments
in the {\bf ab} plane, and antiferromagnetic correlations between the ferromagnetic
clusters, alternatively ordered along the {\bf c} axis [see Fig.\,\ref{Fig1}\,(b,c)]
\cite{FrontzekThesis}. The observed anomalous behaviour in the low-energy optical conductivity spectra below 0.2 eV can be associated with the opening of a pseudogap in the free charge carriers (Drude-type) response. As a result, the vast amount of optical SW of
${\scriptstyle\rm\Delta}N^{\rm D,\bf ab}_{\rm eff}(30\,{\rm
K},\,10\,{\rm K})\approx0.011$ is shifted to the optical band at
0.2\,eV, which we relate to electronic instability of the free charge carriers against the long-range ferromagnetic correlations  in the {\bf ab} plane below $T_{\rm N}$. The estimated amount of optical SW corresponds to a kinetic energy loss of the free charge carriers of about 5\,meV, which correlates well with the characteristic temperature of the broad peak (55 K) in the {\bf c}-axis ac-susceptibility \cite{FrontzekKreyssig06,FrontzekKreyssig07,Paulose,FrontzekThesis}.
This observation can explain the quasi-one-dimensional character this Kondo-lattice
metallic system displays above the N\'eel temperature, and suggests the existence of the chain-type ferromagnetic clusters in this temperature range.

Nevertheless, several features of the observed low-energy
optical conductivity anomalies are difficult to reconcile with a
simple spin-density-wave scenario, where the SW, lost below the gap,
accumulates above the gap. While in the $\bf ab$ plane, the SW changes of the Drude response across $T_{\rm N}$ are nearly compensated by the opposite
changes of the SW of the optical band at 0.2\,eV; and the SW
accumulated by the optical band at 0.6\,eV,
${\scriptstyle\rm\Delta}N^{\rm\bf ab}_{\rm eff}(30\,{\rm
K},\,10\,{\rm K})$\,$\approx$\,0.0036 [see Fig.\,\ref{Fig4}\,(e)], must be
compensated at higher energies, which is an agreement with experimental
observations in Fe-based superconductors and their parent compounds reported
previously \cite{Wang,Schafgans}. This effect, which is accompanied by the low-energy optical conductivity anomalies, requires further experimental and theoretical study.  Additionally, the anomalous increase in the dc conductivity in the $\bf ab$ plane below $T_{\rm N}$ (see the Supplementary Information to this paper) \cite{Majumdar} can be largely accounted for by the reduced scattering rate of the charge
carriers due to freezing out of the spin-flip scattering, caused by
ferromagnetic ordering of the Tb\,$4f$ moments in the formed clusters.
Interestingly, as we have discovered, the intensity of the 0.6\,eV band simultaneously shows an anomalous decrease below $T_{\rm N}$, indicating that the associated electronic excitation is suppressed by the frozen ferromagnetic alignment of the localised magnetic moments in the $\bf ab$ plane [see inset of Fig.\,\ref{Fig3}\,(c)]. We attribute this behaviour to the inter-site low-spin (LS) electronic excitation in an electron pair, where two electrons with antiparallel spins are residing on the same lattice site. An opposite observable trend in the behavior of the 0.2\,eV band suggests that its origin can probaböly be attribute to the inter-site high-spin (HS) electronic excitation. In this case, two electrons with parallel spins are residing on the same lattice site; and the associated electronic excitation is enhanced by the frozen ferromagnetic alignment  of the localised magnetic moments in the $\bf ab$ plane.
According to our earlier optical study of the insulating $d^1$ ($t_{2g}$) system YTiO$_3$ \cite{kovaleva_yto}, the contribution from the inter-site $d_i^1d_j^1$ $\longrightarrow$ $d_i^2d_j^0$ charge excitations to the optical
response leads to four different excited states: a HS $^3T_1$ state at energy
$U^*-3J_H$, two degenerate LS states $^1T_2$ and $^1E$ at energy
$U^*-J_H$, and a LS state $^1A_1$ at energy $U^*+2J_H$ \cite{oles}, where
$U^*$ is the effective Coulomb repulsion of the two electrons with opposite spins on the same orbital, and $J_H$ is the Hund's rule coupling constant. There is an obvious analogy between 5d$^1$ system in Tb$_2$PdSi$_3$ and 3d$^1$ system in YTiO$_3$. Following that, we assign 0.2 eV to the HS-state transition and the 0.6 eV to the first LS state, and estimate the values of $U^*$ at $\approx 0.8$\,eV and $J_H$ at $\approx 0.2$\,eV. Thus, the next LS transition is expected at $\sim$ 1.2 eV, in a good agreement with our data [see
Fig.\,\ref{Fig3}\,(b)].

Having analysed the observed anomalous low-energy optical-conductivity behavior in a Kondo-lattice metal Tb$_2$PdSi$_3$ across $T_{\rm N}$, we discovered that the SW changes of the optical band at 0.2\,eV are in a quantitative agreement with the SW amplitude around 0.2\,eV across the spin-density-wave (SDW) transition at $T_{SDW}=200$ K in the iron arsenide SrFe$_2$As$_2$ \cite{charnukha}.
In this compound, the ground state of the antiferromagnetic SDW instability of a  stripe-type (or collinear) spin configuration was suggested to result
from the nesting between the hole-and-electron Fermi surfaces of itinerant electrons, rather than the superexchange interaction mediated through the of-plane As atoms \cite{Hu}. The origin of the low-energy two-gap behavior, found in optical measurements, and of the higher-energy 0.6\,eV optical feature  is still unclear \cite{Hu}, which makes the nature of the driving force of the SDW instability in the iron arsenide compounds an open and a challenging question. Hence, some close qualitative and quantitative similarities with the optical features discussed in this optical study of the Kondo-lattice
metal Tb$_2$PdSi$_3$ could be of a significant importance.

We note, however, that in the system we  chose to study, the magnetic pattern is more complicated than a regular spin-density wave. Indeed, the long-range magnetic order arising at $T_{\rm N}$ is characterized by different periods along the Tb $3f$ and $1a$ chains, whereas the short-range correlations seen below $T_{\rm N}$ distort even this not particularly regular order. This is an additional evidence that we deal here with quite an irregular structure, most probably formed by spin chains of different lengths, which become frozen at lower temperatures, producing a spin-glass-like state observed in neutron diffraction experiments.

\begin{figure}[b]
\includegraphics[width=8.5cm]{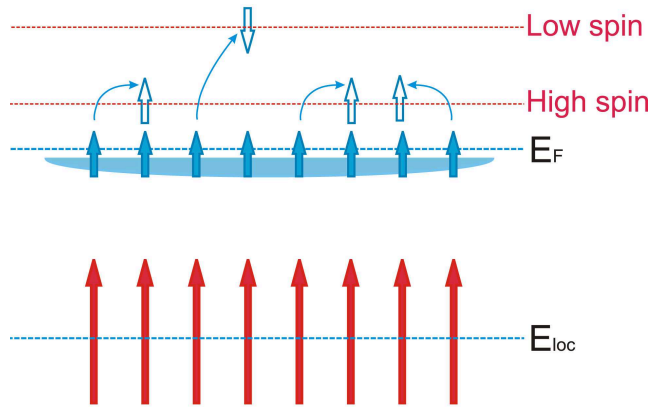}
\caption{{\bf A schematic representation of ferromagnetic metallic cluster
in Tb$_2$PdSi$_3$}. Localisation of itinerant electrons in the string cluster is associated with development of the high-spin (HS) and low-spin (LS) inter-site
electronic transitions. \vspace{-1.5em}}
\label{Fig5}
\end{figure}

We conclude that the observed anomalous low-energy optical-conductivity behavior in a Kondo-lattice metal Tb$_2$PdSi$_3$ occurs due to electron correlation rather than a band structure effect. We propose that magnetism of the large localised Tb 4f magnetic moments  in the metallic ferromagnetic cluster is driven by the electron exchange instability at the Fermi level, similarly to instability of itinerant electrons. The optical SW redistribution involving the free charge carrier Drude resonance (broad and narrow), and the
midinfrared high-spin- (HS)  and low-spin- (LS) state optical bands at 0.2\,eV
and 0.6\,eV respectively, indicates electronic instability of the free charge carriers against coupling in the  metallic ferromagnetic cluster. Thus, by
observing the associated low-energy optical conductivity anomalies, one may elucidate further the evolution in the creation of magnetic clusters.
and estimate the amount of free carriers self-trapped by them in the Kondo-lattice
metal.

In the first instance the magnetic clusters may arise from
magnetic polarons~\cite{Naga,Naga69,Naga72,Kasuya69}, which are formed by
large moments and free carriers. It is well known that
small polarons in general may form string-like objects~\cite{Kus99,Kus00,Kus01}. In the present compound the magnetic polarons may display similar behaviour dictated by a large anisotropy and the quasi-one dimensional character of the magnetic ordering~\cite{FrontzekKreyssig06,FrontzekKreyssig07,Paulose,FrontzekThesis}. Here, the lowest energy of the localised states corresponds to the small
polarons assembled into ferromagnetic string clusters. In a
single cluster, many free particles are self-trapped by the spin
fluctuation involving many large Tb magnetic moments having the spin orientation parallel to the {\bf ab} plane. The spins of the trapped charge carriers and magnetic moments are collinear.

We estimated the size of these ferromagnetic string
clusters, taking into account many-body effects of Coulomb
interaction between electrons, their interaction with localised
large moments, which are ferromagnetically ordered and coupled to each other due to some indirect exchange interaction, induced by itinerant charge carriers. The detailed mechanism of the string clusters formation is discussed in the Supplementary Information, here we only describe its main features. According to this mechanism, the itinerant charge carriers and the localised moments are forming together a potential well, where these charge carriers are self-trapped. Due to such a localisation, the kinetic energy of the trapped electrons increases,
while at the same time their potential energy decreases. As a result, the
total energy of the electrons in the clusters becomes smaller or comparable
with the energy of itinerant charge carriers. The comparison indicates that
a string with many trapped particles and magnetic moments may have lower
total energy (including electronic, magnetic, and Coulomb contributions),
as shown schematically in Fig.\,\ref{Fig5}.

It is important to note that the energies of string clusters having
different lengths are very close to each other. In general, we
expect to find a very broad distribution of such clusters at the
temperature above $T_{\rm N}$ and, possibly, also below this
temperature. Such magnetic clusters give rise to the tail in the
density of states, which appears at the bottom of the conduction
band. The localised states are separated from itinerant ones by a
mobility edge, which is located below Fermi energy.
The broad distribution of string sizes can manifest itself as the broad peak of magnetic ac-susceptibility around $T = 55$ K ~\cite{FrontzekKreyssig06,FrontzekKreyssig07,Paulose,FrontzekThesis}.

Thus, a broad variety of the low-temperature ``exotic'' magnetic phases,
observed in magnetic neutron scattering in an external magnetic field~\cite{FrontzekThesis}
can be associated with the existence of the magnetic clusters as described
above. The clusters can merge into string-like droplets or stripy microdomains, and may be ordered or disordered. At low temperatures, the system exists
in a glassy state, associated with chaotically distributed ferromagnetic microdomains with the short-range order in the basal plane as seen in the magnetic neutron scattering~\cite{FrontzekKreyssig06,FrontzekKreyssig07,Paulose,FrontzekThesis}.
At high temperatures, this state survives until the ferromagnetism vanishes in the vicinity of the critical temperature, associated with the self-localisation
energy of an electron in the cluster. The similar droplets (fluctuons), originally anticipated in the papers by Krivoglaz~\cite{Krivoglaz69,Kriv} and for magnetic systems discussed in~\cite{KaganKuUFN}, can appear as features in the temperature dependence of magnetic susceptibility and resistivity.

\vspace{0.3cm}
\hspace{-1.2em}{\large \bf Methods}\\
\hspace{-1em}{\bf Crystal growth}\\
Careful selection and handling of high-purity  starting materials
and the control of oxygen impurities during the whole preparation
process was essential. The polycrystalline feed rods were prepared
from bulk pieces of rare earth, transition metal, and silicon of
purity 99.9\% or better which were arc-melted several times in a
water-cooled copper crucible under Ar atmosphere to reach
homogenuity. The high-quality single crystals of Tb$_2$PdSi$_3$
were grown by the floating-zone method with optical heating at
melting temperatures $>$ 1500 $^{\circ}$C from near-stoichiometric
polycrystalline feed rods. Post-growth heat treatment was used to
improve the actual  structure of grown crystals. Composition,
microstructure and the perfection of the crystals structure of the
samples were investigated by chemical analysis, optical
metallography and scanning electron microscopy. The
crystallographic orientation of single crystals was determined by
the X-ray Laue back scattering method~\cite{Behr,CrystalGrowth}.

\vspace{0.3cm}
\hspace{-1em}{\bf Experimental approach}\\
The complex dielectric response of Tb$_2$PdSi$_3$ single crystals
was investigated in a wide spectral range using a set of
home-built ellipsometers at Max-Planck-Institut f\"ur
Festk\"orperforschung, Germany. The VIS and UV measurements in the
photon energy range of 0.75--6.0\,eV were performed with a
home-build ellipsometer of rotating-analyzer type, where the angle
of incidence is 70.0$^\circ$  For the temperature-dependent
measurements, the sample was mounted on the cold finger of a
He-flow cryostat with a base pressure of $\sim$\,$2\times10^{-9}$
Torr at ambient temperature.  We used the home-build ellipsometer
with incidence angles ranging from 65$^\circ$ to 85$^\circ$ in
combination with the \textit{Bruker IFS 66v/S} FT-IR spectrometers
for the MIR and NIR measurements in the range of
50\,meV\,--\,1.0\,eV.

\vspace{0.3cm}
\hspace{-1em}

\enlargethispage{2pt}
\vspace{0.3cm}
\hspace{-1.2em}{\large \bf Acknowledgments}\\
We acknowledge discussions with J.~W. Allen, D.~S. Inosov,  A.~V. Boris, A.~Charnukha, D.~Pr\"opper, O.\,V.~Dolgov, D.\,V.~Efremov, L.~Boeri, M.~Frontzek, G.~Jackeli, E.~Kotomin, B.~Keimer and I.~Mazin. We thank A.~N. Yaresko
and V.~N. Antonov for the local-density-approximation calculations. We also thank E.\,V.~Sampathkumaran for providing the resistivity data.

\begin{center}{\large \bf Supplementary Information}

\end{center}

\subsection{DC transport anomalies}

\begin{figure}[h]
        \includegraphics[width=0.7\columnwidth]{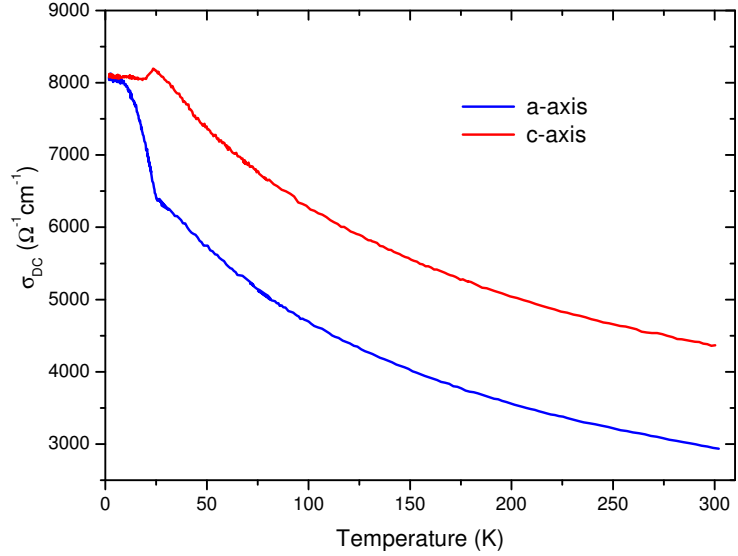}
        \caption{{\bf DC transport anomalies} Temperature dependencies of
        dc conductivity in Tb$_2$PdSi$_3$ single crystal derived from the
        transport measurements by Majumdar
        \textit{et al.} [31].}
        \label{Fig:DC conductivity}
\end{figure}

The dc resistivity measurements in Tb$_2$PdSi$_3$ single crystals [31] revealed clear anomalies below the magnetic transition at $T_{\rm N}=23.6$
K, as illustrated by the corresponding dc conductivity curves in Fig.\,S1\,. While the $\bf a$- and $\bf c$-polarized dc conductivities follow similar trends, with the $\bf c$-axis conductivity higher by $\sim$\,1700 $(\Omega \cdot {\rm cm})^{-1}$ immediately above $T_{\rm N}$, the opposite trends appear below $T_{\rm N}$, providing evidence that magnetic scattering plays
a substantial role in the electronic transport. Surprisingly, the accelerated increase in the $\bf a$-axis conductivity with the simultaneous weak decrease in the $\bf c$-axis conductivity finally results in the isotropic charge transport at low temperatures.

\subsection{Drude-Lorentz analysis of the dielectric function spectra}
To separate the contributions to the complex dielectric response, we performed the classical dispersion analysis by fitting the Drude term and a minimal set of Lorentzian oscillators simultaneously to $\varepsilon_1(\omega)$ and $\varepsilon_2(\omega)$, using a dielectric function of the form [30]
$$\varepsilon(\omega)=\epsilon_{\infty}-\frac{\omega_{\rm p}^2}{\omega^2+{\rm i}\omega \gamma_{\rm D}}+\sum_j\frac{S_j\omega_j^2}{\omega_j^2-\omega^2-{\rm i}\omega\gamma_j}.$$ Here $\omega_{\rm p}$ and $\gamma_{\rm D}$ are the plasma frequency and the damping of free charge carriers, $\omega_j$, $\gamma_j$, and $S_j$ are the peak energy, width, and dimensionless oscillator strength of the $j^{\rm th}$ oscillator, respectively, and $\epsilon_{\infty}$ is the core contribution to the dielectric function. The results
of the dispersion analysis of the {\bf ab}-plane and {\bf c}-axis dielectric function spectra are summarized in Table~S1.\vspace{-1em}

\begin{table*}
\caption{{\bf Drude-Lorentz analysis.}
Parameters of the Drude band and Lorentzian oscillators resulting from the
fit to $\varepsilon_1(\omega,T)$ and $\varepsilon_2(\omega,T)$ in the {\bf ab}-plane and {\bf c}-axis. The values of $\omega_i$ and $\gamma_i$ are expressed
in eV.\smallskip}

\begin{tabular}[c]{@{}l@{~~}l@{~~}D{.}{.}{-1}@{~~}D{.}{.}{-1}@{~~}D{.}{.}{-1}@{~~}D{.}{.}{-1}@{}}
\hline
 &  & \multicolumn{2}{c}{\textbf{ab}-plane\hspace{-2em}~} & \multicolumn{2}{c}{\textbf{c}-axis\hspace{-2.5em}~}\\
 & \hspace{-0.8em}parameters &  \textit{T}\,=\,10\,\textup{K}\hspace{-3.5em}~ & \textit{T}\,=\,30\,\textup{K}\hspace{-3.5em}~ & \textit{T}\,=\,10\,\textup{K}\hspace{-3.5em}~ & \textit{T}\,=\,30\,\textup{K}\hspace{-3.5em}~ \\ \hline
 & \quad$\varepsilon_{\rm inf}$ & 1.4(1) & 1.6(1) & 1.9(1) & 1.9(1) \\
Drude & \quad$\omega_{\rm p}$ & 2.703(5) & 2.750(5) & 2.464(5) &2.498(5)  \\
 & \quad$\gamma_{\rm D}$ & 0.091(2) & 0.098(2) & 0.096(2) & 0.098(2) \\ \hline
 mid-infrared & \quad$\omega_1$  & 0.19(2) & 0.19(2) & 0.20(5) & 0.20(5) \\
bands & \quad$S_{\rm osc}$ & 91(1)\hspace{-1.6em}~ & 87(1)\hspace{-1.6em}~ & 151(7)\hspace{-1.6em}~ & 139(7)\hspace{-1.6em}~  \\
 & \quad$\gamma_1$ & 0.26(5) & 0.26(5) & 0.32(5) & 0.33(5) \\
 & \quad$\omega_2$ & 0.58(2) & 0.59(2) & 0.57(2) & 0.57(2) \\
 & \quad$S_{\rm osc}$ & 21.6(2) & 22.9(2) & 28.1(3) & 27.4(3) \\
 & \quad$\gamma_2$ & 0.74(5) & 0.76(5) & 0.75(5) & 0.75(5) \\ \hline
interband & \quad$\omega_3$ & 1.3(1) & 1.3(1) & 2.2(1) & 2.1(1) \\
 transitions& \quad$S_{\rm osc}$ & 2.7(3) & 2.9(3) & 1.6(2) & 1.6(2) \\
 & \quad$\gamma_3$ & 0.6(1) & 0.6(1) & 1.2(2) & 1.2(2) \\
 & \quad$\omega_4$ & 1.7(1) & 1.7(1) & 2.7(1) & 2.7(1) \\
 & \quad$S_{\rm osc}$ & 4.6(5) & 4.3(5) & 12.8(5) & 12.9(5) \\
 & \quad$\gamma_4$ & 0.9(1) & 0.9(1) & 4.6(3) & 4.7(3) \\
 & \quad$\omega_5$ & 3.4(2) & 3.4(2) & 5.2(5) & 5.2(5) \\
 & \quad$S_{\rm osc}$ & 7.9(5) & 7.6(5) & 0.7(1) & 0.6(1) \\
 & \quad$\gamma_5$ & 4.9(3) & 4.6(3) & 3.1(3) & 3.1(3) \\ \hline
\end{tabular}\vspace{1em}
\end{table*}

\subsection{Magnetic strings and string glass in Kondo-lattice metals: Theory}
Here we show  that  Kondo-lattice metals have a tendency to form large magnetic string clusters. In  a single cluster there are many  polarised spins, which can create a potential well that traps several electrons.
The  magnetic string may be viewed also as a  condensation of magnetic polarons into a linear object. The form of the string is arising due to the long-range Coulomb repulsion, which should be taken into account together with the Hubbard on-site repulsion. Such strings may arise not only in the Kondo-lattice metals
but also in other systems
of narrow band semiconductors.
Their creation is inherent to some kind of instability leading to an electronic phase separation associated with local attraction of small magnetic polarons. The magnetic string clusters may also form a tail in the density of states arising at the bottom of the conduction band, which   can be detected by spectroscopic methods.








Magnetic polarons in antiferromagnetic semiconductors (ferrons) have been introduced  more than 40 years ago [S1].
There, it  was assumed that a single charge carrier has been self-trapped by a ferromagnetic or canted antiferromagnetic region. Such magnetic polaron has been originally named as a ferron and had quite a large size. Of course, the size of the ferron may be large or very small. This depends on the kinetic energy of the trapped electron. Large ferrons have a very simple spherical shape~[S1-S3].
For smaller ferrons, the structures are more complicated [S4].
The central ferromagnetic region of the ferron consists of a magnetized core, which traps the electron. Its surrounding  consists of shells with the oppositely directed moments repelling the trapped electron. The compensating moment of the surrounding oscillates with a period equal to a double lattice constant or larger [S4,S5].
Here, we consider a generalisation of the ferron concept to a multiparticle case, where the central ferromagnetic core traps more than one electron.  The formation of such a state stems primarily from the competition between the energies associated with the Coulomb repulsion between electrons and trapping of the particles by the ferromagnetic core. The minimum energy of the Coulomb repulsion is usually associated with a linear shape of the central trapping region of (see, Refs. [S6,S7]
for details) and therefore such objects have been named electronic strings [S8,S9]
and were detected in oxide materials (see, review article [S10]
and references therein). The formation of strings must be enhanced in the ferromagnetic Kondo-lattice metals since the Coulomb interaction there is partially screened. Therefore, in such materials, we expect to observe ferromagnetic electronic strings, which are described below. Indeed, such evidence obtained
with the use of the optical method is presented in the main body of the paper.

Here we elucidate that an instability of magnetic polarons may give rise
to a creation of a glassy state in the Kondo-lattice metals. That is we have considered small magnetic polarons at zero temperature and found  that the small
polarons become collapsed into a string cluster consisting of many particles trapped by the spin fluctuation and having the same spin orientation. We investigate this instability taking into account
many-body effects of Coulomb interaction for electrons interacting with localised spins ferromagnetically coupled to each other. The long-range Coulomb interaction is taken into account within the Hartree--Fock approximation. We have found that electrons having the spin orientation parallel with respect to the localized background spins are self-trapped into the droplets. The number of the electrons in a single droplet depends on the ratio of the effective Coulomb and the exchange energies.
This state survives until the ferromagnetism vanishes. In the vicinity of the critical temperature the similar droplets (fluctuons) were originally anticipated in the papers by Krivoglaz [S11].

To describe the phenomenon of the droplet or the string formation in the Kondo-lattice metals we start with the Hamiltonian
\begin{equation}
H = t \sum_{ij} c_{i\sigma}^\dagger c_{j\sigma} +  \chi
 \sum_{i\alpha\beta}
c_{i\alpha}^\dagger \vec{\sigma}_{\alpha\beta}
c_{i\beta} \vec{S}_i
+ J \sum_{\left\langle ij \right\rangle } \vec{S}_i \vec{S}_j -\sum_i \vec{S}_i \vec{h}+H_C,
\label{Kondo-Ham}
\end{equation}
where $\vec{\sigma} = (\sigma_x \sigma_y \sigma_z)$ are
Pauli matrices, $t$ is the hopping integral for conduction electrons, which creation and annihilation operators are denoted as $c_{i\sigma}^\dagger$ and $c_{j\sigma}$, respectively. $S_i$ is a Kondo spin localised at the $i$-th site (for simplicity we consider here a cubic lattice) and $h$ is an external magnetic field.
The last term, $H_C,$ is the Hamiltonian of the long-range Coulomb interaction, which is in the Kondo-lattice metals may be screened and replaced by a Hubbard term and a next-neighbour Coulomb repulsion. The constant $\chi$ describes an interaction between free electrons and the Kondo spins. The constant $J$ describes a weak exchange antiferromagnetic interaction between Kondo spins. We
consider the spins in the classical limit, i.e. we treat the spin
operators as classical vectors. This approximation is applicable
only for the systems with large values of localized spins, like $S_0=3/2,...,\infty$.

It is convenient to split our discussion into two parts: at first, we will treat only a single particle interacting with localized moments; and then, in the second part we will take into account the Coulomb interaction in the framework of a Hartree--Fock approximation.

The Schr\"odinger equation describing a single particle interacting with localized magnetic moments has the form
\begin{equation}
-t\Delta\Psi_{\alpha n} + \chi
\vec{S}_n\cdot\vec{\sigma}_{\alpha\beta}
\Psi_{\beta n} = E\Psi_{\alpha n}.
\end{equation}
where the Greek subscripts of the wave function $\Psi_{\alpha\beta}$
describe spin indices, like $\alpha = \pm 1$ and
$\hat{\Delta}$ is a lattice version of the Laplacian operator, which
for the cubic lattice is defined as
\begin{equation}
\hat{\Delta}\Psi_n = -\sum_i(\Psi_n - \Psi_{n+i}),
\end{equation}
where  the  summation is carried out over all the nearest-neighbor sites around the $n$-th site; here, for a simplicity, the spin index $\alpha$ is omitted. With the use of this equation one can construct the total energy of the system, which includes the electronic energy $E$ and the exchange energy corresponding the ferromagnetically localised spins, described by Heisenberg Hamiltonian
$H_H$, i.e.
\begin{equation}
F = E(S_1,...,S_N) + J \sum_{<ij>} \vec{S}_i\vec{S}_j- \sum_i \vec{S}_i \vec{h},
\label{free}
\end{equation}
where we choose $J>0$. As a reference state, we consider a ferromagnetic state formed by localised spins $\vec{S_i} =S_0 \vec{l}$, with $\vec{l} \cdot \vec{l}=1$. Now let us consider
self-trapping of the electrons by a magnetic fluctuation.
If the coupling constant $\chi/J$ is large, the electron
can significantly disturb the spin background. The maximum
value corresponds to the spin flip. Therefore, the spin configuration of a fluctuation at the self-trapping will take the form
\begin{equation}
\vec{S}_{n0} = \left\{ \begin{array} {r}
 -S_0 \vec{l},   \quad  \rm{if}
\quad  1\leq n_x \leq  N  \cr
 \equiv \quad S_0 \vec{l}, \quad \quad \quad
 \quad  \rm{otherwise.}
\end {array}
\right.
\label{sol-sin}
\end{equation}
Note that at large value $\chi/J$ this fluctuation has a
sharp boundary.

After substituting this solution into the Schr\"odinger equation, we get the following system of equations
\begin{equation}
-t\Delta\Psi_{\alpha n} + \chi
\vec{\sigma}_{\alpha\beta}\cdot
 \vec{S}_{n0}   \Psi_{\beta n}  = E\Psi_{\alpha n}.
\label{SE -1}
\end{equation}

It is convenient to choose the system of coordinates with $z$ axis
oriented along the magnetic moments of the ferromagnets, i.e. along $\vec{l}$. In this system of coordinates, these equations for the electrons trapped by the magnetic fluctuation are simplified to
\begin{eqnarray}
\left\{ \begin{array} {r} -t\Delta\varphi_n  = (E+\chi S_0)\varphi_n
 \quad  \rm{if}
\quad  1\leq n_x \leq  N  \cr
 -t\Delta \varphi_n   = (E-\chi S_0)\varphi_n , \quad \quad \quad
 \quad  \rm{otherwise}
\end {array}
\right.
\label{SE-2}
\end{eqnarray}
and analogous equation for electrons with other polarisation $f_n$, i.e. the electron wave function was taken as $\vec{\Psi_n} = (\varphi_n,f_n)$. These equations describe a particle localised in a square and flat potential well of the deepness $-\chi S_0$
and the height $+\chi S_0$. If we assume that this potential
well is infinitely deep and trapped electrons are spin polarised, the solution takes the simple form; the wave function does not vanish
if the index $n$ corresponds to sites within the fluctuation, otherwise it vanishes, i.e. in the limit $\chi/J\rightarrow\infty$ the system of equations \eqref{SE-2} has the following exact solutions: $f_n \equiv 0$ and
\begin{equation}
\varphi_{nx}(k_x)= \left\{ \begin{array} {r}
{{1}\over{\sqrt{N}}} \exp({i k_x n_x}),  \quad  \rm{if}
\quad  1\leq n_x \leq  N  \cr
 \equiv 0, \quad \quad \quad
 \quad  \rm{otherwise.}
\end {array}
\right.
\label{sol-man}
\end{equation}

This single particle wave function corresponds to the following eigenvalue
\begin{equation}
E = -\chi S_0 + Z t-2 t \cos(k_x),
\label{eigenv}
\end{equation}
where $Z$ is a number of nearest neighbours.  Note that the large term $Zt$ associated with the confinement of the electrons may be compensated by a coupling with the ferromagnetic spins of the string $-\chi S_0$.

This solution has a cigar (or string) shape. Such form of the localised state is dictated by the Coulomb interaction, which has  the smallest value for the linear string shape [S6,S7].
The  described magnetic string is  a generalization of  a magnetic ``polaron", which was originally anticipated in the papers by Krivoglaz [S11]
and Nagaev [S12].
Krivoglaz has called ``fluctuon'' an analogous state of a spherical shape created in the vicinity of the critical point of a ferromagnetic phase transition
[S11]
by a cigar shape fluctuation consisting of the flipped magnetic moments is minimal.

Thus, an important role in the formation of the many-particle state  plays the Coulomb interaction which is trying to remove the trapped particles from the trapping well associated with the magnetic fluctuation. Of course, if $M$ particles are trapped by the potential well, they occupy the energy levels of the well according to the Pauli principle, i.e. on the each energy level will be only one particle. On the other hand, in this case, the potential well of such a fluctuation will be created by all $M$ particles and therefore will be $M$ times deeper and wider than in the case of the single particle state.

Thus, in this many particle self-trapping each of $M$ particles having the
up-spin is localised in the potential well self-consistently created by all trapped particles and each particle occupies only a single energy level inside the well. Having these single particles wave functions (Eq. (\ref{sol-man})) we  are now in a position to take into account the long-range Coulomb interaction to apply the Hartree--Fock approximation. With the use of these
single particle wave functions we built up the many-body Hartree--Fock wave function.  The many-body wave function
of the $M$ particles $\Psi(1,2,...,M)$ self-trapped in the string and whose spins are polarised  is taken to be in the form of a Slater
determinant of single particle wave functions, Eq. \eqref{sol-man}
\begin{equation}
\Psi(1,2,...,M)=\frac{1} { \sqrt{M!}} \det\mid\mid \psi_i(k_j)\mid\mid.
\end{equation}
Such a kind of wave functions is proved to be good to describe
the Coulomb correlations in the Hubbard and $t-J$  models [S13].

Then the Coulomb contribution into the total energy of $M$ trapped
particles is estimated as
\begin{eqnarray}
E_c = \langle \Psi \mid \sum_{i<j} \frac{1}{(r_i-r_j)} \mid \Psi\rangle.
\end{eqnarray}
To perform  such an averaging with the use of the Hartree--Fock
many-body wave function, first, we have calculated the pair correlation function. With the aid of the pair correlation function we have calculated the dependence of the Coulomb energy on $N$ and $M$  given by
\begin{equation}
V_{HF}= V \frac{4  M^2}{N} \int_{0}^{\pi}
\frac{dx}{x} \left(1-
\frac{\sin^2(Mx)}{M^2 \sin^2 x} \right),
\label{Hart-Fock}
\end{equation}
where $V = \frac{e^2}{2\bar \epsilon a}$ and the parameter
$\bar{\epsilon}$  is the effective dielectric constant.
The numerical estimation of the integral in Eq. \eqref{Hart-Fock} shows that a function $V_{HF} $ behaves similarly to that obtained independently in the electrostatic approximation, where we  assume that the  $M$ point charges are equidistantly located in the string.  For such an assumption the energy contribution from the long-range part of
the Coulomb interaction is straightforward to calculate. The long-range part of the Coulomb interaction of $M$ particles with
the charge $e$, separated by a distance $a N/M$ and self-trapped into a string of length $N$ is approximately equal to
\begin{equation}
E_C \approx V \left(\frac{M^2\log M}{N}\right).
\label{en-Coulomb}
\end{equation}

The momentum ${\bf k}$ of electrons with polarised spins
is simply quantised if we assume that their wave function satisfies periodic boundary conditions (PBC) along the string. If we employ
other boundary conditions for the trapped electrons (for example, open boundary conditions), the main result will not change drastically.

With the use of PBC the electron momenta along the string are quantised, $k_{nx}=2 \pi n/(aN)$. With the use of Eq.~\eqref{eigenv} and the Pauli exclusion principle, we calculate the expression for the adiabatic potential $J_{N,M}$ describing  $M$ trapped
electrons
\begin{equation}
J_{N,M} = ZtM-\chi S_0 M- 2t \frac{(N-
1)\sin(\pi M/N)}{N\sin(\pi/N)}.
\label{J-spectr}
\end{equation}

If the last two terms in Eq.~\eqref{J-spectr} are larger than the first one,
with the increase of the number of trapped particles,
 $M$,  
the value of the electron energy or the adiabatic potential $J_{N,M}$ decreases. This suggests the possibility of the electronic phase separation in the Kondo-lattice metal and the formation of a large linear object.  However, the size of these objects is limited by the Coulomb interaction $H_C$ (see, the expression for the total Hamiltonian, Eq.(\ref{Kondo-Ham})).

For the next illustration, to estimate the total string energy, $F_S$, we use $E_C$ instead of $V_{HF}$  and add this term to the expression for the adiabatic potential \eqref{J-spectr}.
Because of the negative contribution to $V_{HF}$ arising from
the exchange forces, we gave $V_{HF}\leq E_C$. Such a substitution is, therefore, justified if we are interested in estimating lower limit for the string size due to the Coulomb energy. Also it gives the explicit expression Eq. \eqref{en-Coulomb}, which is convenient for the analysis.

As a result, the total energy $F_S$, which consists of the electron energy $J_{N,M}$ and the energy of the Coulomb repulsion $E_C$, equals
\begin{equation}
F_S  = -\chi S_0 M +Zt M-2t \frac{(N-
1)\sin(\pi M/N)}{N\sin(\pi/N)}+ 
V \left(\frac{M^2\log M}{N}\right).
\label{en-string}
\end{equation}
Here, in the simplest approximation, we neglected the magnetic energy associated with the Heisenberg Hamiltonian $H_H$, related to the last two terms in our Hamiltonian (see Eq. \eqref{Kondo-Ham}). The presented expression for $F_S$
(Eq.~\eqref{en-string}) has an absolute minimum at some fixed number of particles $M$  self-trapped into a string bag having an optimum length, $N$. In insulators, the equations, which determine the optimum values of $M$ and $N$ are obtained by minimization of $F_S$ with respect to $M$ and $N$. In metals, the value of $F_S$ should be compared to the Fermi energy, $E_F$, that is for each value of $N$, the value of $M$ should be determined from the equation $F_S (N,M)=E_F$. In the case of the insulating ferromagnetic string, when $N=M$, we can immediately obtain the equation to determine the number of the trapped electrons, $M$

\begin{eqnarray}
-\chi S_0  +Zt +
 \log M=E_F ,
\label{length--mag-string}
\end{eqnarray}
whence we find
\begin{equation}
M=\exp\left(\frac{E_F+\chi S_0-Z t}{V}\right).
\label{string-length}
\end{equation}
Note that in the Kondo-lattice metals with large Fermi surface
the magnetic strings may have a very large length since the characteristic potential of the Coulomb interaction will be screened
on
large distances and may be very small. Typically, it is described by the the Coulomb potential associated with the next-neighbour interaction.

In general, the strings with the energy lower than the Fermi energy will be also created, and form the Lifshiz tail in the density of states which may be spread well in the bandgap.

The comparison of the string energy $F_S$ per electron (see Eq.~\eqref{en-string}) with the  Fermi energy indicates that a string of the length $N$ with $M$ trapped particles may have a total energy (including  electronic, magnetic, and Coulomb contributions), which is lower than  the Fermi energy or the total energy of $M$ separated
magnetic polarons [S1].
Therefore, the state associated with $M$ magnetic polarons (either separated, forming a Wigner crystal, or forming a Fermi liquid) is unstable and will collapse into a mixture of different strings forming a complicated
magnetic order.

The described linear strings will also have much lower energy than  sheet configurations, like a single circular or square spot (droplet),
 consisting of $M$ sites inside the droplet with $M$ self-trapped particles. It is clear, for example, from the simple electrostatic arguments presented  that for the  configuration of a single circular droplet a contribution to the total energy from the Coulomb interaction is strongly increased compared to a string configuration, while a contribution from the exchange interaction being short-ranged
remains the same. Although these droplets have higher total energy than strings per particle, it is still smaller than the energy of a  single small polaron.

Thus,  we arrive at the conclusion that in Kondo-lattice metals {\bf small magnetic polarons  are unstable; this instability induces the formation of strings, which may be ordered or form highly disordered glassy state}.
The similar state was discussed earlier for the case of pyrochlore manganites
[S14].
These strings are created by an exchange interaction between free particles and localised Kondo magnetic moments.  The string may not only have a linear form but may be also bent, curved or even have a shape of a closed loop.  Such curved configurations will probably correspond to low-energy excitations of the string.

The concept of the string arising in Kondo-lattice metals introduced here is very general. Depending on the fermion filling of the strings, the individual strings may be both insulator and metallic, which are coexisting with each other. Such a large variety of the different types of strings may give rise to different novel effects which could arise in materials with narrow bands: both semiconductors and metals. In general, for any narrow band Kondo-lattice metal, we expect the formation of a new type of the electronic order formed by the long linear objects -- strings. There, the criterion for the string formation is significantly improved, because in the Kondo-lattice metals the total energy of the strings is comparable or lower than the Fermi energy, from one side, and  the Coulomb repulsion is significantly screened, from the other side. It seems that for such metals it is a common phenomenon that there is a coexistence of strings and free fermions, which balance is dictated by an interplay
of Fermi energy, Coulomb and exchange forces. These complicated issues will be discussed in our future publications.


In summary, we demonstrate here that Kondo-lattice metals represent an example of a new electronic state of matter where there is a coexistence of the magnetic clusters with free fermions. The magnetic moments associated with individual clusters  may be disordered at high temperatures and ordered in a some kind of antiferromagnetic order at low temperatures.
The antiferromagnetic-type order is associated with dipole character of magnetic interaction between the  total magnetic moments of individual clusters. In general, we expect that the low-temperature state has a glassy character.
The application of an external field increases the length of the
strings and therefore increases a size of ferromagnetic domains, hence, diminishing the spin fluctuation scattering in the sample.

{\it Acknowledgements.}  We are very grateful to E. I. Rashba for illuminating discussions.


\noindent\subsection{Supplementary references}

\noindent[S1]~
Nagaev, E. L. {\it Pisma v Zh. Eksp. Teor. Fiz.} {\bf 6,} 484 (1967). [{\it
JETP Lett.} {\bf 6,} 18 (1967)]; {\it Zh. Eksp. Teor. Fiz.} {\bf 54,} 228 (1968). [{\it Sov. Phys. JETP} {\bf 27,} 122 (1968).]

\noindent[S2]~
Mott, N. and Zinamon, Z. {\it Rep. Prog. Phys.} {\bf 33,} 81 (1970).

\noindent[S3]~
Nagaev, E. L. {\it Phys. Reports} {\bf 346,} 388 (2001).

\noindent[S4]~
Nagaev, E. L. {\it Pisma v Zh. Eksp. Teor. Fiz.} {\bf 74,} 472 (2001). [{\it
JETP Lett.} {\bf 74}, 431 (2001).]

\noindent[S5]~
Nagaev, E. L. {\it Phys. Rev. B} {\bf 66,} 104431 (2002).

\noindent[S6]~
Kusmartsev, F. V. {\it J. de Physique IV} {\bf 9,} Pr10-321 (1999).

\noindent[S7]~
Kusmartsev, F. V. {\it Phys. Rev. Lett.} {\bf 84,} 530 (2000); {\it Phys. Rev. Lett.} {\bf 84,} 5026 (2000).

\noindent[S8]~
Kusmartsev, F. V., Di Castro, D., Bianconi,
G. \&  Bianconi, A. {\it Phys. Lett. A} {\bf 275,} 118 (2000).

\noindent[S9]~
Kusmartsev,  F. V. {\it Int. J. Mod. Phys. B} {\bf 14,} 3530 (2000).

\noindent[S10]~
Kusmartsev, F. V. {\it Contemp. Phys.} {\bf 45,} 237 (2004).

\noindent[S11]~
Krivoglaz, M. A. {\it Usp. Fiz. Nauk} {\bf 113,} 617 (1973). [{\it Sov. Phys. Uspekhi} {\bf 16,} 856 (1974).]

\noindent[S12]~
Nagaev, E. L. {\it Physics of Magnetic Semiconductors}
(Mir Publishers, Moscow, 1983).

\noindent[S13]~
Kusmartsev, F. V. {\it Phys. Rev. B} {\bf 43,} 6132 (1991).

\noindent[S14]~
Kusmartsev, F. V. {\it Physica B} {\bf 284-288,} 1422 (2000).








%
%
%
%




\end{document}